\documentclass[11pt,twoside]{article}

\usepackage{asp2006}
\usepackage{epsf}
\usepackage{psfig}
\usepackage{lscape}

\begin{document}

\title{Ultraluminous Infrared Galaxies at $1.5<z<3$ occupy dark matter haloes of mass $\sim6\times10^{13}M_{\sun}$}   

\author{D. Farrah\altaffilmark{1}, C. J. Lonsdale\altaffilmark{2,6}, C. Borys\altaffilmark{3}, F. Fang\altaffilmark{2}, I. Waddington\altaffilmark{4}, S. Oliver\altaffilmark{4}, M. Rowan-Robinson\altaffilmark{5}, T. Babbedge\altaffilmark{5}, D. Shupe\altaffilmark{2}, M. Polletta\altaffilmark{6}, H. E. Smith\altaffilmark{6}, J. Surace\altaffilmark{2}}

\altaffiltext{1}{Department of Astronomy, Cornell University, Ithaca, NY 14853, USA}
\altaffiltext{2}{IPAC, California Institute of Technology, Pasadena, CA 91125, USA}
\altaffiltext{3}{Department of Astronomy, University of Toronto, Toronto, Canada}
\altaffiltext{4}{Astronomy Center, University of Sussex, Falmer, Brighton, UK}
\altaffiltext{5}{Astrophysics Group, Imperial College, London SW7 2BW, UK}
\altaffiltext{6}{CASS, University of California at San Diego, La Jolla, CA 92093, USA}

\begin{abstract} 
We present measurements of the spatial clustering of ultraluminous infrared galaxies in two redshift intervals, $1.5<z<2.0$ and $2<z<3$. 
Both samples cluster strongly, with $r_{0}=14.40\pm1.99 h^{-1}$Mpc for the $2<z<3$ sample, and $r_{0}=9.40\pm2.24 h^{-1}$Mpc for the 
$1.5<z<2.0$ sample, making them among the most biased galaxies at these epochs. These clustering amplitudes are consistent with both 
populations residing in dark matter haloes with masses of $\sim6\times10^{13}M_{\sun}$. We infer that a minimum dark matter halo mass 
is an important factor for all forms of luminous, obscured activity in galaxies at $z>1$. Adopting plausible models for the growth of DM 
haloes with redshift, then the haloes hosting the  $2<z<3$ sample will likely host the richest clusters of galaxies at z=0, whereas 
the haloes hosting the $1.5<z<2.0$ sample will likely host poor to rich clusters at z=0.
\end{abstract}

\section{Introduction}   
When we look at the night sky, we see that galaxies seem to be arranged in a particular way. One might expect that galaxies 
would be distributed randomly, much as grains of sand would if you threw a handful across the floor, but instead, they seem to 
trace elegant structures; galaxy clusters are connected to each other by long filaments, interspersed with large voids, where 
few or no galaxies are seen. The drivers behind the formation of these 'large-scale structures' have been the subject 
of intense study and debate for over thirty years; how did the Universe go from being smooth and homogeneous just after the 
Big Bang to the clumpy, clustered Universe we see today? 

At the core of current theories for the formation of these structures is the premise that the evolution of the total mass 
distribution is described by the gravitational collapse of primordial density fluctuations, and that this evolution is traced 
by the evolution of galaxies. Overdense regions, or `haloes', are predicted to undergo mergers to build haloes of increasing mass, 
with galaxies forming from the baryonic matter in these haloes. This framework of `biased' hierarchical buildup \citep{col,hatt}
has proven to be remarkably successful in explaining several important aspects of galaxy and large-scale structure formation.

This paradigm is however not without its problems. A good example of these problems concerns the evolution of massive 
($\geq10^{11}M_{\sun}$) galaxies. We might expect that massive galaxies form slowly, with many halo mergers needed to build 
up the required large baryon reservoirs, and indeed some galaxies do appear to form in this way \citep{van,bel}. There is 
however evidence that many massive galaxies may form at high redshift \citep{dun,blak} and on short timescales \citep{ell}, 
directly counter to early, `naive' model predictions. Intriguingly, recent surveys \citep[e.g.][]{eal,scot} have uncovered a huge population of distant, IR bright 
sources that are plausible candidates for being rapidly forming, massive galaxies, however they are so numerous 
that even recent models have difficulty in explaining their counts, and invoke a wide variety of solutions (e.g. \citet{bau}). 

It seems likely therefore that there are strong, but subtle links between distant, IR/sub-mm bright galaxies, and the formation of 
large-scale structures.  Therefore, we need observations that relate the properties of these galaxies with the underlying dark 
matter distribution. Motivated by this, we have used data from the Spitzer Wide Area Infrared Extragalactic Survey 
(SWIRE, \citet{lon}) to select large samples of distant Ultraluminous Infrared Galaxies (ULIRGs, L$_{ir}\geq 10^{12}$L$_{\sun}$) 
and study their clustering evolution with redshift. We assume $H_{0}=100h$ km s$^{-1}$ Mpc$^{-1}$, $\Omega=1$, and $\Omega_{\Lambda}=0.7$. 
The results presented here were originally published in \citet{far06a,far06b}.

\section{Analysis and results} 
To select high redshift ULIRGs, we use the 1.6$\mu$m emission feature, which arises due to photospheric emission 
from evolved stars. When this feature is redshifted into one of the IRAC channels then that channel exhibits a `bump' \citep{sim,saw}. 
A complete discussion of the source selection and characterization methods is given in Lonsdale et al 2006 (in preparation), which we summarize here. 

Our sources are taken from the ELAIS N1 and ELAIS N2 fields, and the Lockman Hole, covering 20.9 square degrees in total. We first selected those sources 
fainter than R=22 (Vega), and brighter than 400$\mu$Jy at 24$\mu$m. Within this set, we selected two samples that displayed a 
`bump' in the 4.5$\mu$m and 5.8$\mu$m channels, i.e. where $f_{3.6}<f_{4.5}>f_{5.8}>f_{8.0}$ for one sample (the `B2' sample) and where 
$f_{3.6}<f_{4.5}<f_{5.8}>f_{8.0}$ for the other sample (the `B3' sample). This resulted in a total of 1689 B2 sources and 1223 B3 sources. 

For both samples we used {\sc Hyper-z} \citep{bol} to estimate redshifts, the results from 
which place most of the B2 sources within $1.5<z<2.0$, and most of the B3 sources within $2.2<z<2.8$. From the best fits we also 
derived IR luminosities and power sources; the requirement that the sources have $f_{24}>400\mu$Jy demands an IR luminosity of 
$\geq10^{12}$L$_{\odot}$ for all the sources, with (most objects having) a starburst as the dominant power source, with star formation 
rates of $\geq200$M$_{\odot}$yr$^{-1}$. Similarly, the presence of the 1.6$\mu$m feature demands a 
minimum mass of evolved stars of $\sim10^{11}$M$_{\odot}$. Both the B2 and B3 sources are thus good candidates for being 
moderately massive galaxies harboring an intense, obscured starburst, making them similar in nature to both local ULIRGs, and 
high-redshift SMGs. 

We measured the angular clustering of both samples using the methods described in \citet{far06a,far06b}, which we summarize here. We found that the levels of 
angular clustering seen in the three fields were consistent with each other to within 0.5$\sigma$, and so combined the angular 
clustering measures for each sample over the three fields. To quantify the strength of clustering, we fit both datasets with a 
power law, $\omega(\theta) = A_{\omega}\theta^{1-\gamma}$, where $\gamma=1.8$ and $A_{\omega}$ is the clustering amplitude; 
$A_{\omega}=0.0125\pm0.0017$ for the B3s and $A_{\omega}=0.0046\pm0.0011$ for the B2s. To convert these angular clustering amplitudes to spatial 
clustering amplitudes, we invert Limbers equation: 

\begin{equation} \label{equ:spatc3}
\frac{r_{0}(z)}{f(z)}=\left[\frac{H_{0}^{-1}A_{\omega}cC\left[\int_{a}^{b}\frac{dN}{dz}\,dz\right]^{2}}{\int_{a}^{b}\left(\frac{dN}{dz}\right)^{2}E(z)
D_{\theta}^{1-\gamma}(z)f(z)(1+z)dz}\right]^{\frac{1}{\gamma}}
\end{equation}

\noindent where $f(z)$ parametrizes the redshift evolution of $r_{0}$, and:

\begin{equation} \label{equ:spatc4}
C = \frac{\Gamma(\gamma/2)}{\Gamma(1/2)\Gamma([\gamma-1]/2)}, E(z) = [\Omega_{m}(1+z)^{3} + \Omega_{\Lambda}]^{\frac{1}{2}}
\end{equation}

We derived $dN/dz$ from the photometric redshift distributions \citep{lon3}. For the 
B2 sources this is a Gaussian centered at $z=1.7$ with a FWHM of $1.0$, and for the B3 sources this is a Gaussian centered at 
$z=2.5$ with a FWHM of $1.2$. The resulting correlation lengths are $r_{0}=9.4\pm2.24h^{-1}$Mpc for the B2 sources, and 
$r_{0}=14.4\pm1.99h^{-1}$Mpc for the B3 sources.

\section{Discussion} 
To place these clustering results in context, we consider two models. The first parametrizes the spatial correlation 
function, $\xi$, as a single power law in comoving coordinates, where the comoving correlation length, 
$r_{0}(z)$, is:

\begin{equation} \label{equ:spatc2}
r_{0}(z) = r_{0}f(z), f(z)=(1+z)^{\gamma-(3+\epsilon)}
\end{equation}

\noindent Here the choice of $\epsilon$ determines the redshift evolution \citep{phl,ove}. Several cases are usually quoted. First is `comoving clustering', 
where haloes expand with the Universe, and $\epsilon=\gamma-3$; in this case clustering remains constant. Second 
is the family of models for which $\epsilon\geq0$, for which clustering increases with time. Examples of this family 
include (a) `stable' clustering, for which $\epsilon\simeq0$ (in this case the haloes are frozen in {\it proper} coordinates,  
(b) the predicted evolution of clustering of the overall dark matter distribution, where $\epsilon\simeq\gamma-1$ \citep{car2b}, 
and $r_{0}\simeq5$ at z=0 \citep{jen}, (c) `linear' clustering, where $\epsilon=1.0$. A cautionary note to this is that detailed interpretations of 
clustering evolution from these models suffer from several theoretical flaws \citep{mos,smi}, and so should be thought of as 
qualitative indicators rather than quantitative predictions. We therefore simply use the `stable' and `linear' models as 
indicators of the possible range of halo clustering amplitude with redshift. 

\begin{figure}
\plotone{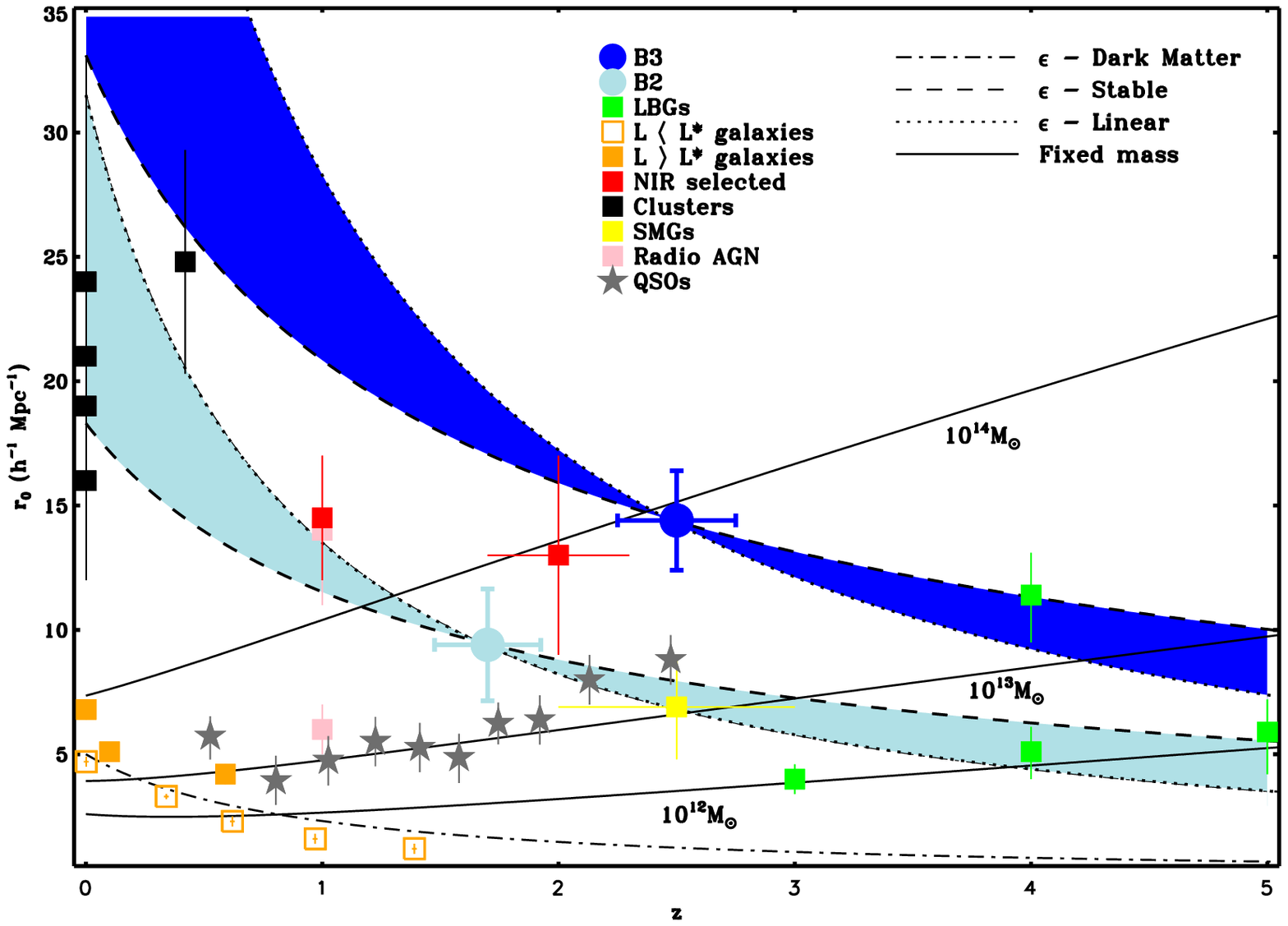}
\caption{Comoving correlation length, $r_{0}$, vs. redshift. Other data are taken from \citealt{mos,ove,dad,bla,ouc,ade,cro2,geo,all}. 
The `Fixed mass' lines show the predicted clustering amplitude of haloes of a given mass at any particular redshift, whereas 
the $\epsilon$ lines show the predicted clustering amplitude of an individual halo for three halo growth models, described 
in the text. The `Stable' and `Linear' lines give a qualitative indicator of the range of how DM haloes may grow with redshift, 
and we have normalized `Stable' and `Linear' lines to the clustering amplitudes of the B2s and the B3's. The shaded regions therefore 
indicate what these haloes may host at lower and higher redshifts - the haloes hosting B3s may contain an optically bright LBG at 
$z\simeq4$ (upper green point), and grow to host very rich galaxy clusters at z=0, whereas the haloes hosting B2 sources may 
contain optically fainter LBGs at $4<z<5$, SMGs at $z\sim2.5$, radio-bright AGN (upper pink triangle) and (old) EROs at $z\simeq1$, and poor 
to rich clusters at z=0.  
\label{r0plot}}
\end{figure}

The second class of model comprises those in which comoving correlation lengths increase with {\it increasing} redshift. These 
models introduce `bias', $b(z)$, between the galaxies and the underlying dark matter. An example of such models are the `fixed 
mass' models \citep{mata,mos}, which predict the clustering strength of haloes of a specified mass at any given redshift. In Figure 
\ref{r0plot} we plot the $\epsilon$ model for dark matter, `stable'and `linear' epsilon models normalized to the B2 and B3 
clustering strengths, the fixed halo mass models for halo masses of $10^{12}$M$_{\odot}$, $10^{13}$M$_{\odot}$, and $10^{14}$M$_{\odot}$, 
the $r_{0}$ values for the B2 and B3 galaxies, and the spatial correlation lengths of other galaxy populations taken from the literature. 

With the the uncertainties described earlier firmly in mind, we use Figure \ref{r0plot} to explore the 
relationships between our samples, the underlying dark matter, and other galaxies. Both the B2s and the B3s are strongly clustered, with correlation lengths much 
higher than that predicted for the overall DM distribution. Both B2s and B3s cluster significantly more strongly than 
optical QSOs at their respective epochs, and B3s cluster more strongly than SMGs. Based on the \citet{mata} models, then we derive 
{\it approximate} 1$\sigma$ halo mass ranges of $10^{13.7}<$M$_{\odot}<10^{14.1}$ for the B3s, and $10^{13.5}<$M$_{\odot}<10^{13.9}$ 
for the B2s. Interestingly, halo masses comparable to these were recently derived for an independent sample of high-redshift ULIRGs by \citet{mag}.  

The most interesting comparison is however between the two samples themselves. The clustering evolution of QSOs with redshift \citep{cro2} 
may mean that there is a `minimum' host halo mass for QSO activity, below which no QSO is seen, of $\sim5\times10^{12}$M$_{\odot}$. The 
correlation lengths for the B2 and B3 samples are consistent with the same conclusion but for a $\sim6\times10^{13}M_{\odot}$ `minimum' DM 
halo mass. Taken together, these results imply that a minimum halo mass is a `threshold' factor for {\it all} forms of luminous activity in 
galaxies, both starbursts and AGN. It is also interesting to speculate on what the host haloes of B2 and B3 sources contain at lower and 
higher redshifts. We might expect that a halo hosting a B3 source could contain an optically bright LBG at $z\sim4$, followed by a B3 at 
$z\sim2.5$, possibly accompanied by other (near-IR selected) star forming systems \citep{dad,dad2}, before evolving to host a rich galaxy 
cluster at low redshifts.  The occupants of a halo hosting a B2 galaxy would however probably be different. We would expect that such a halo 
could contain an SMG at $z\sim2.5$, and optically fainter LBGs at $4<z<5$ (though probably not LBGs at $z\sim3$). At lower redshifts such a 
halo might host a radio-bright AGN and or ERO at $z\sim1$, and a (poor to rich) cluster at $z=0$. We conclude that ULIRGs at $z\geq1.5$ as 
a class likely signpost stellar buildup in galaxies in clusters at $z=0$, with higher redshift ULIRGs signposting stellar buildup in 
galaxies that will reside in more massive clusters at lower redshifts.

\acknowledgements
Support for this work, part of the Spitzer Space Telescope Legacy Science Program, was provided by NASA through an award issued by JPL 
under NASA contract 1407.

\end{document}